\documentclass{fizik}%
\usepackage{multicol,epsfig}

\vol{30}
\pyear{2006}
\received{18.07.2006}

\author[BULUTAY]{
\textbf{Ceyhun BULUTAY}\\
\textit{Department of Physics, Bilkent University, 06800, Ankara-TURKEY}\\
}

\title{Pseudopotential-based full zone $k \cdot p$ technique\\ 
for indirect bandgap semiconductors: \\
Si, Ge, diamond and SiC}

\begin{document}
\maketitle

\begin{abstract}
The $k \cdot p$ is a versatile technique that describes the semiconductor 
band structure in the vicinity of the bandgap. The technique can be extended to full 
Brillouin zone by including more coupled bands into consideration. For completeness, a 
detailed formulation is 
provided where the associated $k \cdot p$ parameters are extracted from the local empirical 
pseudopotential method in the form of band edge energies and generalized momentum matrix elements. We 
demonstrate the systematic improvement of the technique with the proper choice of the band edge states 
for the group-IV indirect bandgap semiconductors: Si, Ge, diamond and SiC of the 3C cubic phase. 
The full zone agreement is observed to span an energy window of more than 20~eV for Si, and 40~eV 
for the diamond with the 15-band pseudopotential-based $k \cdot p$ approach.
\end{abstract}

\section{Introduction}
For the electronic and optical processes in semiconductors involving states near 
the bandgap, the $k \cdot p$ technique has been the first resort for many 
researchers~\cite{chuang}. The $k \cdot p$ method was introduced by Bardeen~\cite{bardeen}
and Seitz~\cite{seitz}. Luttinger and Kohn extended the technique to degenerate 
bands to govern the valence band edge structure of common group-IV elemental and 
III-V compound semiconductors as described by a $3\times 3$ matrix~\cite{lk}.
In the case of direct bandgap semiconductors, appending the 
lowest conduction band as well as including the spin-orbit split-off within the valence 
band lead to a $4\times 4$ matrix which is also known as the Kane's model~\cite{kane}. 
Soon afterwards Pidgeon and Brown extended this to $8\times 8$ matrix by including 
the spin for all bands in the presence of an external magnetic field~\cite{pb}.
These measures have enlarged the number of coupled bands and also improved the 
overall agreement of the band effective masses with the experimental values. On the 
other hand, their validity spanned only about 15\% of the full Brillouin 
zone~\cite{richard} which can be for instance sufficient for the description of not too small
excitons. However, they fall short for applications involving the larger 
portion of the Brillouin zone such as the optical absorption spectra or the high-field 
transport characteristics. The classical work in the direction to extend the multiband $k \cdot p$ to 
full zone is due to Cardona and Pollak~\cite{cp} which resulted in a $15\times 15$ 
$k \cdot p$ Hamiltonian. After this period of progress over the technique, researchers 
especially for device physics applications have routinely employed the 4-band and the 
8-band variants of $k \cdot p$ together with the envelope function approximation for 
incorporating the confinement effects~\cite{bastard}. Wood and Zunger have termed this 
combination as the ``standard model" and they have compared it with the empirical 
pseudopotential method (EPM)~\cite{wood}. Both qualitative and significant quantitative 
errors were identified 
mainly due to $L-$ and $X-$derived states which were not accounted by the 8-band 
$k \cdot p$ approach~\cite{wood}. The particular concerns on the use of the envelope 
functions were relieved by Foreman~\cite{foreman} as well as by Burt who 
also clarified the correct operator orderings~\cite{burt}.

For the nanotechnology applications involving quantum dots and nanocrystals, the device 
sizes are reduced in all three dimensions which results in the sampling of the {\it full} 
Brillouin zone. Once again there is the necessity to extend the multiband $k \cdot p$ 
to full zone for technologically-important semiconductors. The increase of bands also 
results in the proliferation of the so-called Luttinger-like parameters which cannot be 
easily extracted from experiments as was the fortunate case for 3-bands where 
effective masses were directly related~\cite{chuang}. One possibility is to make use of the 
tight-binding model as done by Jancu {\it et al.} who have systematically determined the
14-band $k \cdot p$ parameters from the 40-band tight-binding band structure~\cite{jancu}. 
Another route is to resort to the classical Cardona-Pollak methodology~\cite{cp} where 
the empty lattice plane wave basis is used; Richard {\it et al.} followed this approach 
which lead to 30-bands taking into account the spin-orbit coupling. An even simpler 
alternative is to make use of the already existing EPM
band structure data to generate the multiband $k \cdot p$ parameters~\cite{wang96}. It is 
this last choice that we pursue in this work to obtain the multiband $k \cdot p$ band 
structure for group~IV indirect bandgap semiconductors: Si, Ge, diamond and SiC 
of the 3C cubic form. Our basic motivation behind this choice is that these 
semiconductors act as embedded core materials for nanocrystals where the confinement 
effects alleviate the poor luminescence properties of their bulk 
counterparts~\cite{brus}. However, the bulk electronic structure is nevertheless
inhereted to low-dimensions, with the further complication from the interface states. 
We believe that for certain nanotechnology applications, full zone $k \cdot p$ band 
structure that can easily be extracted from EPM as demonstrated in the following sections 
may still be a suitable choice compared to more demanding atomistic approaches.

\section{Theory}
The bulk crystal Hamiltonian described by a local one-electron effective crystal 
potential, $V_{\mbox{\small{xtal}}}(\vec{r})$ is given by
\begin{equation}
\label{hamiltonian}
\left[ \frac{p^2}{2m_0}+V_{\mbox{\small{xtal}}}(\vec{r})\right]\psi_{n\vec{k}}(\vec{r})=
E_n(\vec{k})\psi_{n\vec{k}}(\vec{r})\, ,
\end{equation}
where $\vec{p}=-i\hbar\nabla$, $m_0$ is the free-electron mass, $n$ is the band 
index, and $\vec{k}$ is the crystal wavevector. Inserting for the Bloch wave functions 
$\psi_{n\vec{k}}(\vec{r})=\frac{1}{\sqrt{V}}e^{i\vec{k}\cdot\vec{r}}u_{n\vec{k}}(\vec{r})$, 
where $V$ is the crystal volume, we obtain the well-known the $k \cdot p$ expression
\begin{equation}
\label{kp}
\left[H_0+\frac{\hbar}{m_0}\vec{k}\cdot\vec{p}\right]u_{n\vec{k}}(\vec{r})=\left[ E_n(\vec{k})-
\frac{\hbar^2 k^2}{2m_0} \right] u_{n\vec{k}}(\vec{r})\, ,
\end{equation}
where $H_0=\frac{\hbar^2 k^2}{2m_0}+V_{\mbox{\small{xtal}}}(\vec{r})$. Note that the cell-periodic functions 
$\lbrace u_{n\vec{k}}(\vec{r}) \rbrace $
satisfy a Sturm-Liouville type eigenvalue equation with the eigenfunctions being complete when all 
(i.e., infinite) bands are included at a fixed wavevector $\vec{k}$, say $\vec{k}_0$. 
Based on this completeness at a chosen $\vec{k}_0$ we can expand an arbitrary cell-periodic function, 
$u_{j\vec{k}}(\vec{r})$ as
\begin{equation}
\label{expand1}
u_{j\vec{k}}(\vec{r})=\sum_{m=1}^{N_b}b_{j\vec{k},m\vec{k}_0}u_{m\vec{k}_0}(\vec{r})  \, . 
\end{equation}
Inserting this form into the bulk Hamiltonian given by Eq.~(\ref{hamiltonian}) leads to the 
following eigenvalue equation that becomes exact as $N_b\to\infty$:
\begin{equation}
\label{eigen}
\sum_{m=1}^{N_b} \left[ H_{n\vec{k}_0,m\vec{k}_0}(\vec{k})-\delta_{nm}E_j(\vec{k}) \right] 
b_{j\vec{k},m\vec{k}_0}=0\/,  \, n \in [1,N_b]\, , 
\end{equation}
where
\begin{equation}
\label{eqa}
H_{n\vec{k}_0,m\vec{k}_0}(\vec{k})\equiv \left[ E_m(\vec{k}_0)+
\frac{\hbar^2}{2m_0}\left( k^2-k_0^2\right)\right] \delta_{nm}+
\frac{\hbar}{m_0}\left( \vec{k}-\vec{k}_0\right)\cdot\vec{p}_{n\vec{k}_0,m\vec{k}_0}\, ,
\end{equation}
and 
\begin{equation}
\label{mommat}
\vec{p}_{n\vec{k}_0,m\vec{k}_0}\equiv \langle u_{n\vec{k}_0}\vert\vec{p}\vert u_{m\vec{k}_0}\rangle\, ,
\end{equation}
corresponds to the momentum matrix element.

For computations, obviously one cannot include infinite bands, so it is more viable to build up an 
expansion set from a number of band and wavevector combinations chosen with a physical insight 
mainly reflecting the band extrema states. In other words, rather than using a fixed 
$\vec{k}_0$ we form an expansion over various band and wavevector indices, $\{ m,\vec{k}_i \}$ as
\begin{equation}
\label{overlap}
u_{j\vec{k}}(\vec{r})=\sum_{\{ m,\vec{k}_i \}} b_{j\vec{k}, m\vec{k}_i} u_{m\vec{k}_i}(\vec{r})\, .
\end{equation}
Since the cell-periodic functions are not necessarily orthonormal at different wave vectors, 
we define their overlap as
\begin{equation}
\label{Delta}
\Delta_{{n\vec{k}_s},{m\vec{k}_i}}\equiv\langle u_{n\vec{k}_s}\vert u_{m\vec{k}_i}\rangle\, .
\end{equation}
Hence, the expansion coefficients $b_{j\vec{k}, m\vec{k}_i}$ satisfy the following equation
\begin{equation}
\label{eqb}
\sum_{\{ m,\vec{k}_i \}} \left \{\left[ E_{m}(\vec{k}_i)-\frac{\hbar^2k_i^2}{2m_0}\right ]+
\frac{\hbar}{m_0}\left( \vec{k}-\vec{k}_i\right)\cdot\vec{p}-\left[ E_{j}(\vec{k})-
\frac{\hbar^2k^2}{2m_0} \right ] \right\} b_{j\vec{k}, m\vec{k}_i} u_{m\vec{k}_i}(\vec{r})=0 \, .
\end{equation}
Projecting to $\vert u_{n\vec{k}_s}\rangle$ results in a generalized eigenvalue equation for the 
expansion coefficients 
\begin{equation}
\label{eqc}
\sum_{\{ m,\vec{k}_i \}} \left[ H_{n\vec{k}_s,m\vec{k}_i}(\vec{k})-
\Delta_{{n\vec{k}_s},{m\vec{k}_i}}E_{j}(\vec{k})
\right ]  b_{j\vec{k}, m\vec{k}_i}=0 \, ,\, \forall n,\vec{k}_s \, ,
\end{equation}
where $\{ n,\vec{k}_s \}$ also belong to the same chosen expansion set as $\{ m,\vec{k}_i \}$; here
\begin{equation}
\label{eqd}
H_{n\vec{k}_s,m\vec{k}_i}(\vec{k})\equiv \left[ E_{m}(\vec{k}_i)+ \frac{\hbar^2}{2m_0}\left( k^2-k_i^2\right)\right ] 
\Delta_{{n\vec{k}_s},{m\vec{k}_i}}+\frac{\hbar}{m_0}\left( \vec{k}-\vec{k}_i\right)
\cdot\vec{p}_{n\vec{k}_s,m\vec{k}_i}\, ,
\end{equation}
with
\begin{equation}
\label{genmommat}
\vec{p}_{n\vec{k}_s,m\vec{k}_i}\equiv \langle u_{n\vec{k}_s}\vert\vec{p}\vert u_{m\vec{k}_i}\rangle\, ,
\end{equation}
corresponding to the generalized momentum matrix element linking indirect transitions. 

In summary, one solves Eq.~(\ref{eqc}) for the desired band $j$ and the wavevector 
$\vec{k}$ with the only ingredients 
being the band energies, $E_{m}(\vec{k}_i)$ and the generalized momentum matrix elements, 
$\vec{p}_{n\vec{k}_s,m\vec{k}_i}$. These can easily be extracted from a local pseudopotential 
approach; the cell-periodic functions can be expressed in the Fourier representation as
\begin{equation}
\label{overlap2}
u_{m\vec{k}_i}(\vec{r})=\frac{1}{\sqrt{\Omega_0}}\sum_{\vec{G}}B_{m\vec{k}_i}
({\vec{G}})\, e^{i{\vec{G}}\cdot{\vec{r}}} \, .
\end{equation}
where $\{ \vec{G} \}$ correspond to reciprocal lattice vectors and $\Omega_0$ is the volume of the 
primitive cell. Hence the overlap (Eq.~(\ref{Delta})) 
and the generalized momentum matrix elements (Eq.~(\ref{genmommat})) are given in terms of Fourier 
coefficients $B_{m\vec{k}_i}({\vec{G}})$ of the cell-periodic functions as
\begin{equation}
\label{Delta2}
\Delta_{{n\vec{k}_s},{m\vec{k}_i}}= \sum_{\vec{G}}B_{n\vec{k}_s}^*({\vec{G}})\, 
B_{m\vec{k}_i}({\vec{G}}) \, ,
\end{equation}
and 
\begin{equation}
\label{genmommat2}
\vec{p}_{n\vec{k}_s,m\vec{k}_i}= \sum_{\vec{G}}B_{n\vec{k}_s}^*({\vec{G}})\, \hbar\vec{G}\, 
B_{m\vec{k}_i}({\vec{G}})\, ,
\end{equation}
respectively.

\begin{figure}[h]
\begin{center}
\includegraphics[scale=1]{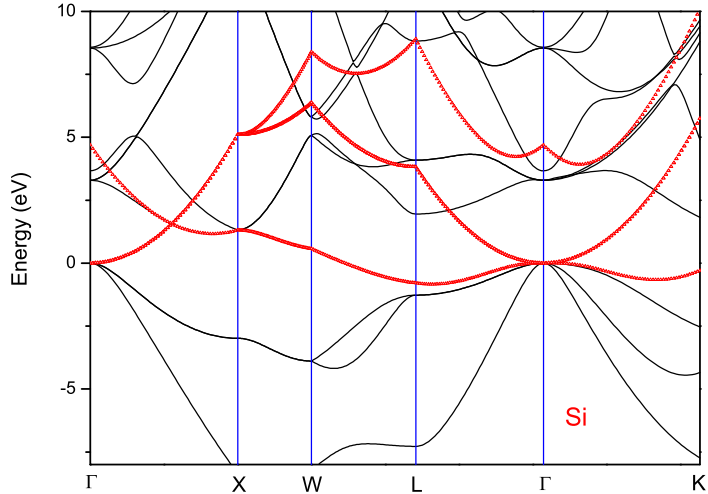}
\end{center}
\caption{(Originals in color) EPM (black lines) vs. 4-band $k \cdot p$ (red symbols) band structure for Si.}
\label{fig1}
\end{figure}

\section{Results}
We apply this basic idea to group~IV indirect bandgap semiconductors: Si, Ge, diamond and SiC 
of the 3C cubic phase. Their local EPM form factors are taken from Refs.~\cite{wang94,saravia} 
for Si and Ge, respectively and from Ref.~\cite{pennington} for the diamond and SiC. 
Spin-orbit coupling effects are neglected which can actually become significant for the case 
of Ge~\cite{saravia}.
Another important technical remark is about the EPM cut off energies: we observed that 
even though the EPM band energies (i.e., eigenvalues) converge reasonably well with cut off energies 
as low as 5-10~Ry, the corresponding Bloch functions (i.e., eigenvectors) require substantially 
higher values to converge. The results to follow are obtained using 14, 16, 33, and 22~Ry for 
Si, Ge, diamond and SiC, respectively.

\begin{figure}[p]
\begin{center}
\includegraphics[scale=1]{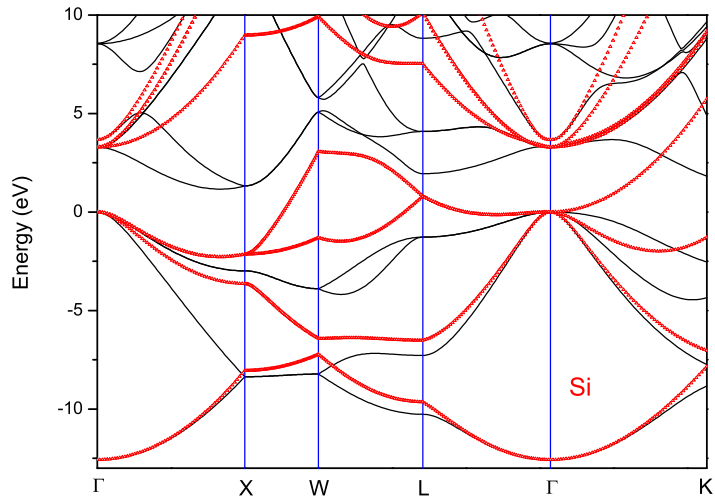}
\end{center}
\caption{(Originals in color) EPM (black lines) vs. 8-band $k \cdot p$ (red symbols) band structure for Si.}
\label{fig2}
\end{figure}

\begin{figure}[p]
\begin{center}
\includegraphics[scale=1]{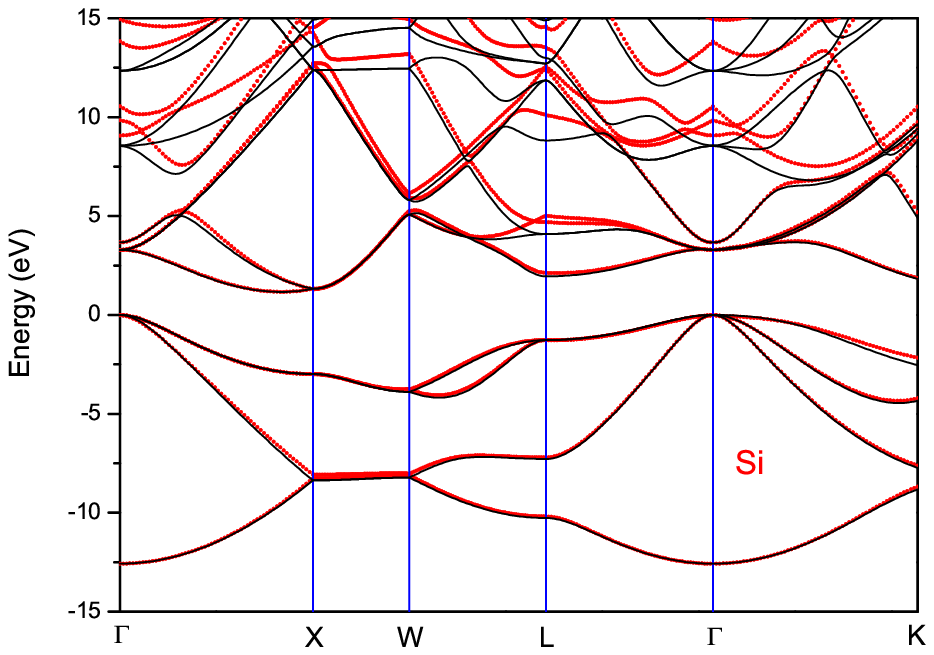}
\end{center}
\caption{(Originals in color) EPM (black lines) vs. 15-band $k \cdot p$ (red symbols) band structure for Si.}
\label{fig3}
\end{figure}

Using silicon as a testbed, we demonstrate how to choose and improve the $k \cdot p$ basis set. 
First, we start with the four-band $k \cdot p$ which is known to yield reasonable results for direct 
bandgap semiconductors like GaAs. These four states are taken to be the highest three valence 
band (VB) states at the $\Gamma$ point and the lowest conduction band (CB) state at one of 
the $X$ points, (1,0,0). The resultant band structure displayed in 
Figure~\ref{fig1} indicates that apart from the four band edge energies even the band curvatures turn 
out to be wrong. To improve this situation, the eight-band $k \cdot p$ is generated by employing 
the highest four VB and the lowest four CB states, all from the $\Gamma$ point. The agreement 
with the EPM band structure shown in Figure~\ref{fig2} is once again not acceptable due to 
remarkable deviation away from the $\Gamma$ point. Also note that the heavy hole band acquires 
the wrong curvature along both the $\Gamma-L$ and  $\Gamma-K$ directions. To remedy these 
shortcomings, we form a set by including from the $\Gamma$ point the band indices 1 to 8 
(as in the previous case), from the $X$ point 
bands 3 to 6, and from the $L$ point bands 3 and 4, and finally from the $K$ point the fifth band 
making altogether 15 states. Here the band index 4 corresponds to highest VB, and 5 to lowest CB.
The band structure of this pseudopotential-based 15-band $k \cdot p$ approach is shown in 
Figure~\ref{fig3} where excellent agreement is observed over an energy window of about 20~eV.

\begin{figure}[p]
\begin{center}
\includegraphics[scale=1]{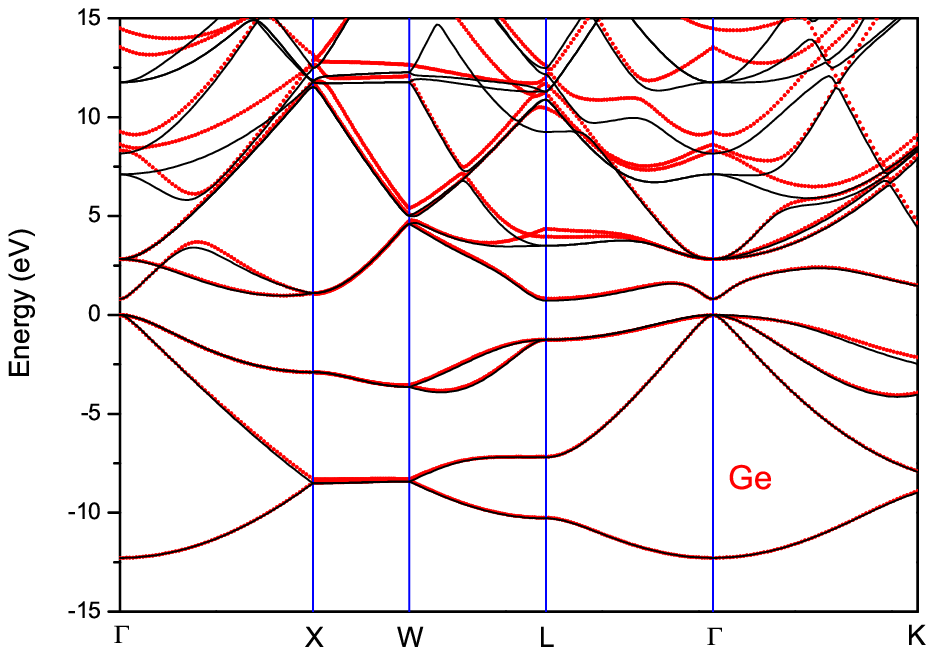}
\end{center}
\caption{(Originals in color) EPM (black lines) vs. 15-band $k \cdot p$ (red symbols) band structure for Ge.}
\label{fig4}
\end{figure}

\begin{figure}[p]
\begin{center}
\includegraphics[scale=1]{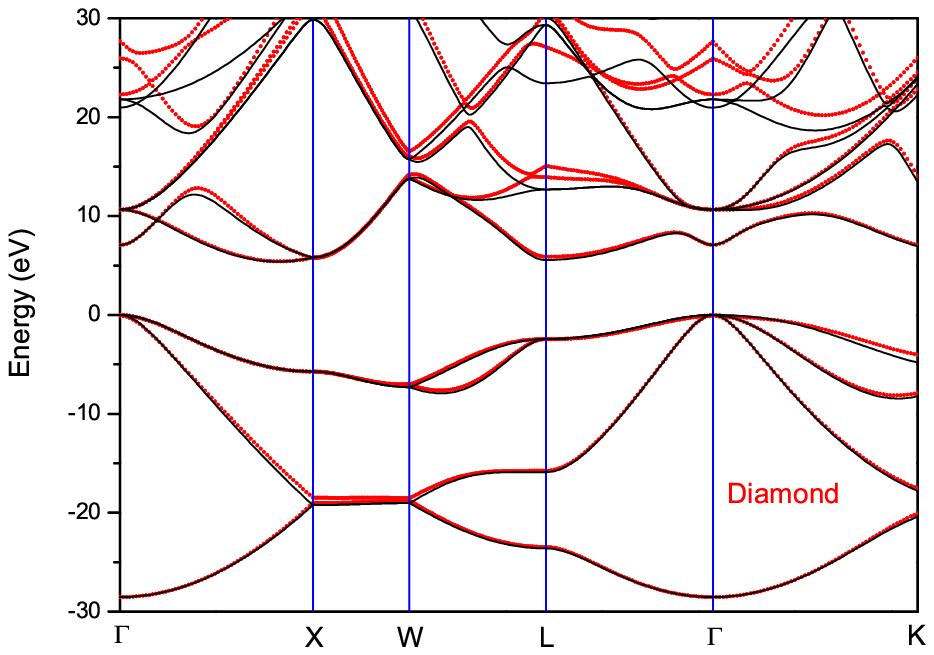}
\end{center}
\caption{(Originals in color) EPM (black lines) vs. 15-band $k \cdot p$ (red symbols) band structure for diamond.}
\label{fig5}
\end{figure}

\begin{figure}[h]
\begin{center}
\includegraphics[scale=1]{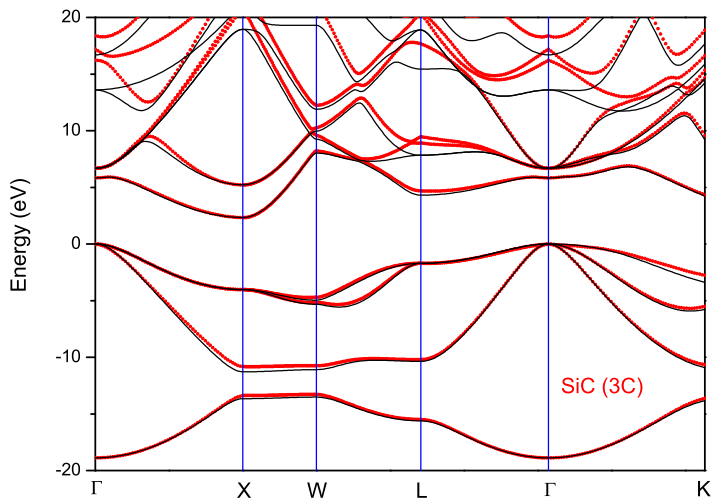}
\end{center}
\caption{(Originals in color) EPM (black lines) vs. 15-band $k \cdot p$ (red symbols) band structure for SiC 
in 3C cubic phase.}
\label{fig6}
\end{figure}

Using the same states for the 15-band $k \cdot p$ approach for the other indirect bandgap 
semiconductors Ge, diamond and SiC, we reach the same performance with Si as displayed in 
Figs.~\ref{fig4}-\ref{fig6}, respectively. For the diamond which is a wide bandgap 
semiconductor, the full zone agreement is observed to span an energy window of more than 40~eV.

\section{Conclusions}
The multiband $k \cdot p$ technique can be systematically improved by including more bands. 
This is particularly needed for indirect bandgap semiconductors where 4- and 8-band approaches fail. 
The band coupling parameters which are in the form of generalized momentum matrix elements and the 
associated band edge energies for the selected states can be extracted from EPM. The technique can 
also benefit from {\it ab initio} local pseudopotential band structure provided that the bandgap shrinkage 
due to local density approximation is corrected~\cite{martin}.
We believe that for certain nanotechnology applications, full zone $k \cdot p$ band 
structure that can easily be generated with the inputs from EPM as demonstrated in this work 
may become the suitable choice compared to more demanding atomistic approaches.
Therefore, what remains to be done is to check the performance of the proposed 15-band 
$k \cdot p$ framework on several low-dimensional applications and to see whether the 
``farsightedness'' of the conventional $k \cdot p$ approach with the small 
number of bands \cite{zunger} is cured.

\section*{Acknowledgments} 
I would like to congratulate Prof. Dr. M. Tomak, my Ph.D. former supervisor, for 
his $60^{th}$ birthday and acknowledge his lasting inspiration with deep respect.
This work has been supported by the European FP6 Project SEMINANO with the 
contract number NMP4 CT2004 505285 and by the Turkish Scientific and Technical Council 
T\"UB\.ITAK within COST 288 Action.

\end{document}